\documentclass[aps,prx,twocolumn,amsmath,amssymb,notitlepage,floatfix, superscriptaddress, longbibliography]{revtex4-2}

\usepackage{color}
\usepackage{graphicx}
\usepackage{cancel}
\usepackage{amsfonts}
\usepackage{amssymb}
\usepackage{amsthm}
\usepackage{url}
\usepackage{nicefrac}

\usepackage{siunitx}
\usepackage{hyperref}
\usepackage{float}
\usepackage{multirow}  % multirow tabular
\usepackage{dcolumn}   % Align table columns on decimal point
\usepackage{bm}        % bold math
\usepackage{stmaryrd}  % for mapsfrom symbol
\usepackage{enumerate}
\usepackage{stmaryrd}
\usepackage{ulem}
\usepackage{siunitx}
\usepackage{braket}
\usepackage{dsfont}

\usepackage[dvipsnames]{xcolor}
\hypersetup{
	colorlinks,
	linkcolor={green!50!black},
	citecolor={blue!50!black},
	urlcolor={blue!80!black},
	citecolor   = {red!50!black} %Colour of citations
}

\renewcommand{\vec}[1]{\boldsymbol{\mathbf{#1}}}

\begin{document}

\title{Boosting the Edelstein effect of two-dimensional electron gases by ferromagnetic exchange}
%\title{Boosted Edelstein effect in ferromagnetic two-dimensional electron gases}

\author{Gabriel Lazrak}
\affiliation{Unit\'e Mixte de Physique, CNRS, Thales, Universit\'e Paris-Saclay, 91767, Palaiseau, France}

\author{B\"orge G\"obel}
\affiliation{Institut f\"ur Physik, Martin-Luther-Universit\"at, Halle-Wittenberg, 06099 Halle (Saale), Germany}

\author{Agnès Barthélémy}
\affiliation{Unit\'e Mixte de Physique, CNRS, Thales, Universit\'e Paris-Saclay, 91767, Palaiseau, France}

\author{Ingrid Mertig}
\affiliation{Institut f\"ur Physik, Martin-Luther-Universit\"at, Halle-Wittenberg, 06099 Halle (Saale), Germany}

\author{Annika Johansson}
 \email{annika.johansson@mpi-halle.mpg.de}
\affiliation{Max Planck Institute of Microstructure Physics, Weinberg 2, 06120 Halle (Saale), Germany}

\author{Manuel Bibes}
 \email{manuel.bibes@cnrs-thales.fr}
\affiliation{Unit\'e Mixte de Physique, CNRS, Thales, Universit\'e Paris-Saclay, 91767, Palaiseau, France}

\date{\today}% It is always \today, today,
             %  but any date may be explicitly specified

\begin{abstract}
Strontium titanate (SrTiO$_3$) two-dimensional electron gases (2DEGs) have broken spatial inversion symmetry and possess a finite Rashba spin-orbit coupling. This enables the interconversion of charge and spin currents through the direct and inverse Edelstein effects, with record efficiencies at low temperature, but more modest effects at room temperature. Here, we show that making these 2DEGs ferromagnetic enhances the conversion efficiency by nearly one order of magnitude. Starting from the experimental band structure of non-magnetic SrTiO$_3$ 2DEGs, we mimic magnetic exchange coupling by introducing an out-of-plane Zeeman term in a tight-binding model. We then calculate the band structure and spin textures for increasing internal magnetic fields and compute the Edelstein effect using a semiclassical Boltzmann approach. We find that the conversion efficiency first increases strongly with increasing magnetic field, then shows a maximum and finally decreases.
% Original: This field dependence is caused by the competing effects of out-of-plane spin polarization and enhanced splitting of band pairs, induced by the out-of-plane magnetic field.
This field dependence is caused by the competition of the exchange coupling with the effective Rashba interaction. While enhancing the splitting of band pairs amplifies the Edelstein effect, weakening the in-plane Rashba-type spin texture reduces it.
\end{abstract}

%\keywords{Suggested keywords}%Use showkeys class option if keyword
                              %display desired
                              
\date{\textcolor{RoyalBlue}{\today}}
\maketitle

%\tableofcontents

\section{Introduction}
Since the discovery of a quasi-two-dimensional electron gas (2DEG) at the interface between the two band insulators LaAlO$_3$ and SrTiO$_3$ (STO) \cite{ohtomo_high-mobility_2004}, 2DEGs at oxide surfaces and interfaces have attracted a lot of attention due to their very rich physics. Superconductivity \cite{reyren_superconducting_2007}, magnetism \cite{brinkman_magnetic_2007, li_coexistence_2011}, gate tunable metal-insulator and superconductor-insulator transitions \cite{thiel_tunable_2006, caviglia_electric_2008} as well as Rashba spin-orbit coupling \cite{caviglia_tunable_2010} make them also promising for applications \cite{cen_oxide_2009, kornblum_conductive_2019}. Additionally, recent experiments have revealed their unprecedented efficiency for spin to charge current interconversion \cite{lesne_highly_2016, Vaz2019}, which is key for new devices such as the magneto-electric spin transistor proposed by Intel for beyond CMOS computing schemes \cite{manipatruni_scalable_2019}. As described in more detail in Refs.~\cite{lesne_highly_2016, Vaz2019}, in those 2DEGs, the inversion symmetry breaking at the interface results in a built-in electric field perpendicular to the interface and thus an additional effective term in the Hamiltonian of the system, the Rashba term, that lifts the spin degeneracy and locks the spin and momentum degrees of freedom. In the simplest case of linear Rashba coupling, this results in two slightly different Fermi contours with opposite spin chiralities. In the inverse Edelstein effect, the injection of a pure spin current perpendicular to the interface can be interpreted as a shift of the two contours in opposite directions to accommodate the spin accumulation. This slight non-equivalence leads to the generation of a charge current within the 2DEG (spin-charge conversion). Its reciprocal effect, the direct Edelstein effect, is achieved by applying an electric field that leads to a reoccupation of states, often visualized as a shift of both Fermi contours in the same direction. This results in a net spin density generation that can diffuse as a pure spin current into an adjacent material (charge-spin conversion).

Recently, possibilities to further enlarge or control the functionalities of STO 2DEGs have emerged. One approach relies on making the 2DEG ferroelectric by exploiting the large electric-field- or Ca-doping-induced ferroelectric character in STO \cite{noel_non-volatile_2020, brehin_switchable_2020}, thereby enabling a non-volatile control of the transport and spin-charge interconversion properties. Another is by inducing spin polarization in the 2DEG by depositing a magnetic layer on top of STO \cite{de_luca_transport_2014, gunkel_defect_2016, stornaiuolo_tunable_2016, zhang_magnetic_2017, kormondy_large_2018, gan_diluted_2019, di_capua_orbital_2022}. Combining both strategies, multiferroic 2DEGs have been realized \cite{brehin_coexistence_2023}, opening an avenue for ferroelectrically controllable chiral spin textures in 2DEGs and providing a new playground for non-volatile spin–orbitronics and non-reciprocal physics.

Here, by combining a tight-binding Hamiltonian and a semiclassical Boltzmann approach, we predict that introducing ferromagnetism in STO 2DEGs can enhance the spin-charge interconversion efficiency by nearly one order of magnitude. The non-monotonic dependence of this efficiency with the amplitude of the induced magnetization is explained by the interplay between magnetic exchange coupling and Rashba-like spin-orbit coupling, that causes a sub-band splitting, as well as a magnetically induced out-of-plane spin polarization. This enhancement is of great promise to obtain spin-charge interconversion large enough for room temperature applications.

\section{Band structure and spin texture}
\begin{figure*}%[t!]
    \centering
    \includegraphics[width=\textwidth]{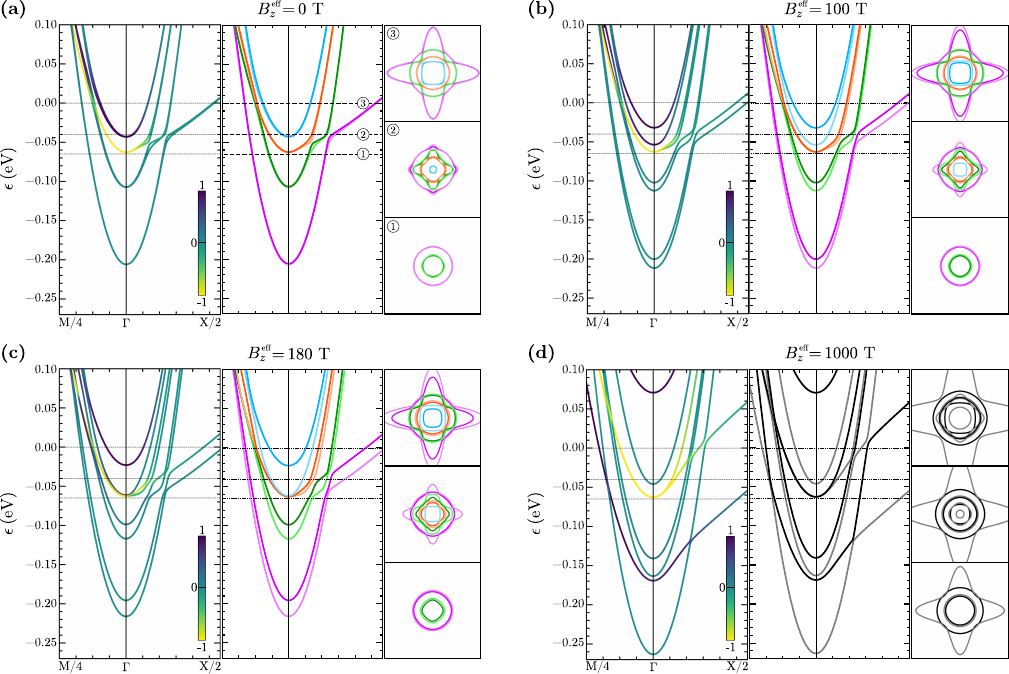}
    \caption{\textbf{Band structure and iso-energy lines of the 2DEG at STO interfaces}.
    The band structure is computed using the tight-binding model introduced in the Appendix~\ref{sec:model_Hamiltonian}, for four exchange field strengths ($B^\text{eff}_z$ = $\SI{0}{\tesla}$ (a), $\SI{100}{\tesla}$ (b), $\SI{180}{\tesla}$ (c), and $\SI{1000}{\tesla}$ (d)). The middle graph of each panel displays the band structure using four colors (magenta, green, orange, blue), with the lower energy bands shown in lighter shades and the higher energy bands shown in darker shades for each pair. On the right, the Fermi lines (with $|k_x|, |k_y| \leq \pi/2a$) illustrate the splitting of the bands for three specific energies ($\SI{0}{\milli\electronvolt}$, $\SI{-40}{\milli\electronvolt}$, and $\SI{-65}{\milli\electronvolt}$ in that order). %The spin and orbital character of each sub-band is depicted on the left. The colorbar represents the projection of the eigenstates onto states with either parallel or antiparallel out-of-plane spin and orbital moments: `1' corresponds to parallel, `-1' to antiparallel, and `0' is the state with zero orbital moment (only spin). 
    The expectation value of the operator $\mathcal{L}_z  \mathcal{S}_z$ (normalized to $\hbar^2/2$) is depicted on the left.
    For panel (d), the band structure and Fermi contours are displayed in black and gray due to significant intertwining of the bands, making it unreasonable to group them by pairs.
}
    \label{fig1}
\end{figure*}
We use an effective eight-band tight-binding model introduced in Refs.~\cite{khalsa_theory_2013, Zhong2013, Vivek2017, Vaz2019} in order to model the $t_{2g}$ electronic states which are relevant for the formation of the 2DEG at STO interfaces. Our model includes two $d_{xy}$ orbitals as well as one $d_{yz}$ and one $d_{zx}$ orbital and has been demonstrated to appropriately approximate the electronic structure of the 2DEG at the STO surface, as well as the interface with LaAlO$_3$ and AlO$_x$~\cite{Vivek2017, Vaz2019}. For details of this model, we refer to Refs.~\cite{Vaz2019, johansson_spin_2021} and to Appendix~\ref{sec:model_Hamiltonian}. In addition to atomic spin-orbit coupling, the broken inversion symmetry at the interface allows for an inter-atomic orbital mixing term, which leads to an effective Rashba term causing a Rashba-like spin splitting of the bands~\cite{khalsa_theory_2013, Zhong2013}. In order to simulate ferromagnetism originating from an adjacent ferromagnetic layer, we introduce an additional magnetic exchange coupling term to the Hamiltonian, $\mathcal H = \mathcal H_\text{STO} + \mathcal H_\text{ex}$, with $\mathcal H_\text{STO}$ the Hamiltonian of the unperturbed STO interface, and  the exchange coupling
\begin{equation}
\label{eq:exchange_Hamiltonian} 
\mathcal H_\text{ex} = -\frac{J_\text{ex}}{\hbar} \left( g_l \vec{\mathcal L} + g_s \vec{\mathcal S} \right) \cdot \hat{\vec M } 
\end{equation}
Here, $\hbar$ is the reduced Planck constant, $\hat{\vec M} $ is the direction of the magnetization,  $J_\text{ex}$ quantifies the exchange coupling between the conduction electrons' orbital/spin moments and the magnetization, $\vec{\mathcal L}$ and $\vec{\mathcal S}$ are the operators of the orbital angular momentum and the spin, respectively, with  $g_l$ and $g_s$ the corresponding Landé factors. The Hamiltonian~\eqref{eq:exchange_Hamiltonian} has the form of a Zeeman Hamiltonian~\cite{Manchon2009}
\begin{equation}
    \label{eq:Zeeman}
   \mathcal H_\text{ex} \widehat{=} \mathcal H_\text{Z}= \frac{\mu_\text B}{\hbar} \left( g_l \vec{\mathcal  L }+ g_s \vec{\mathcal  S} \right) \cdot \vec B^\text{eff}
\end{equation}
with $\mu_\text B$ the Bohr magneton and $\vec B^\text{eff}= - J_\text{ex} \hat{\vec M} / \mu_\text B$ an effective magnetic field originating from the finite magnetization and acting on the electronic states. The Landé factors are assumed $g_l=1$ and $g_s=2$, following Ref.~\cite{johansson_spin_2021}. A representation of the spin and orbital moment operators in the basis of the relevant $t_{2g}$ orbitals can be found in Eqs.~\eqref{eq:L_operator}-\eqref{eq:S_operator}.

In the following, the influence of magnetic exchange coupling on the band structure as well as the charge-spin conversion efficiency are discussed in terms of the effective magnetic field $\vec B^\text{eff}$. Importantly, this field, originating from magnetic exchange interaction, can induce energy splitting in the band structure of a few $\SI{10}{\milli\electronvolt}$ \cite{Krempasky2016}, which corresponds to a $B^\text{eff}$ of a few hundred Tesla. Therefore, the magnetic field $B^\text{eff}$ discussed in the following is much larger than external magnetic fields which can be applied experimentally but could correspond to the field induced by a magnetic exchange interaction in the 2DEG.

First, we examine the electronic band structure of the 2DEG at STO interfaces under the influence of an out-of-plane magnetic exchange field, using the model Hamiltonian introduced in Eq.~\eqref{eq:exchange_Hamiltonian}. 
Figure~\ref{fig1} illustrates the band structure at four different out-of-plane exchange field strengths ($B^\text{eff}_z$ = $\SI{0}{\tesla}$, $\SI{100}{\tesla}$, $\SI{180}{\tesla}$, and $\SI{1000}{\tesla}$) within an energy range of $\SIrange{-270}{100}{\milli\electronvolt}$. For each exchange field strength, the band structure as well as iso-energy lines at three selected energies ($\SI{-65}{\milli\electronvolt}$, $\SI{-40}{\milli\electronvolt}$, and $\SI{0}{\milli\electronvolt}$) are shown to illustrate the influence of the exchange field on the band structure. 

At zero field (Figure~\ref{fig1}a), the splitting of each band pair (marked magenta, green, orange, and blue in the right panel) is solely due to the atomic spin-orbit coupling  and antisymmetric hopping (called orbital mixing in Refs.~\cite{Vivek2017, Vaz2019}), which lift the twofold spin degeneracy. Close to the band edge of each band pair, the band structure is isotropic with circular iso-energy lines (\textcircled{\raisebox{-0.9pt}{1}} in the figure). The heavy bands' Fermi contours take the form of two perpendicular ellipses. The maximum splittings along $\Gamma$X are observed in the region from $\SIrange{-65}{-40}{\milli\electronvolt}$, where avoided crossings occur, the first one between the green and orange band pairs and the second one between the magenta and green band pairs, as highlighted by the Fermi surface at $\SI{-40}{\milli\electronvolt}$ (\textcircled{\raisebox{-0.9pt}{2}} in the figure). In these regions, we observe a strong deviation from the simple Rashba model for free electrons.

Upon increasing the exchange field strength, we observe the expected linear increase of Zeeman-like splitting for each band pair, with the notable exception of the orange band pair that remains unsplit at $\Gamma$. 
To understand this band-dependent splitting caused by the exchange field, it is crucial to analyze the spin and orbital composition of the bands. By breaking the inversion symmetry at the (001) interface, the $t_{2g}$ bands become inequivalent in energy which is why the $d_{xy}$ bands appear at a lower energy near the $\Gamma$ point compared to the $d_{yz}$ and $d_{xz}$ bands. Due to SOC, states that are close in energy hybridize. In our case, the $d_{yz}$ and $d_{xz}$ states form the superpositions $d_{m_l=\pm 1}=(-id_{yz}\mp d_{xz})/\sqrt{2}$, meaning that the cubic atomic orbitals $d_{yz}$, $d_{xz}$ form the atomic orbitals $d_{m_l=\pm 1}$. They are characterized by the quantum numbers $l=2, m_l=\pm 1$, with $l$ the orbital angular momentum quantum number, and $m_l$ the quantum number of the out-of-plane orbital angular momentum operator.

The magnetic exchange field couples with both orbital angular momentum and spin, see Eq.~\eqref{eq:exchange_Hamiltonian}. States with purely $d_{xy}$, $d_{yz}$, or $d_{zx}$ character exhibit zero orbital angular momentum. However, since the $t_{2g}$ orbitals hybridize they may possess nonzero expectation values of the out-of-plane orbital angular momentum $L_z=\hbar m_l$, because $m_l\neq 0$ as explained above. The left panel of each subfigure of Fig.~\ref{fig1} illustrates the spin and orbital character of the electronic states. %Here, we show the projection of the eigenstates to states with either parallel or antiparallel expectation values of the out-of-plane orbital ($m_l$) and spin ($m_s$) angular momenta. 
Here, we show the expectation value of the product of out-of-plane orbital and spin operators, $\mathcal L_z \mathcal S_z$.
On the colorbar: `1' means that the corresponding eigenvalues have the same sign (e.g. $m_l=1$, $m_s=\nicefrac{1}{2}$, the parallel state), `-1' means opposite sign (e.g. $m_l=1$, $m_s=-\nicefrac{1}{2}$, the antiparallel state) and `0' means zero out-of-plane orbital quantum number (e.g. $m_l=0$).

In the third lowest band pair (orange) spin and orbital angular momenta are antiparallel because the eigenstates are superpositions of states with opposite signs of the quantum numbers $m_l$ and $m_s$ ($d_{m_l=-1,m_s=1/2}$ and $d_{m_l=1,m_s=-1/2}$). In the fourth band pair (blue), they are parallel  (superpositions of states with the same sign of $m_l$ and $m_s$: $d_{m_l=1,m_s=1/2}$ and $d_{m_l=-1,m_s=-1/2}$). When a magnetic field is applied, the bands are polarized with respect to spin and orbital momenta. Due to the quantum numbers discussed above, the effects of the orbital- and spin-induced band splitting are compensated for the orange band pair $\Delta\epsilon = 2\mu_B (g_l \Delta L_z + g_s \Delta S_z)/\hbar\approx 0$ because $\Delta L_z \approx -2 \Delta S_z$ and $g_s=2g_l$, but are enhanced in the blue band pair $\Delta\epsilon = 2\mu_B (g_l \Delta L_z + g_s \Delta S_z)/\hbar\approx 4\mu_B|\vec{B}^\text{eff}|$.

Since the lower two band pairs (magenta and green) consist of almost purely $d_{xy}$ states at the $\Gamma$ point at low magnetic fields, the orbital angular momentum $L_z$ is suppressed for these bands which is why they only split up due to the spin contribution: $\Delta\epsilon =2\mu_B g_s \Delta S_z/\hbar\approx 2\mu_B|\vec{B}^\text{eff}|$. These bands could only experience a considerable orbital polarization if they hybridized (a) with the orange and blue bands or (b) with $d_{x^2-y^2}$ states to form the complex orbitals $d_{m_l=\pm 2}=(d_{x^2-y^2}\pm id_{xy})/\sqrt{2}$ that are characterized by $m_l=\pm 2$.

At specific field strengths, the Zeeman-like splitting causes band crossings at $\Gamma$. This leads to increased band mixing, making it unreasonable to discuss them as pairs. At $\SI{180}{\tesla}$, the first such crossing occurs between the orange bands and lower blue band. We will explore the implications of this crossing on the spin-charge interconversion efficiency later. Thus, the center panel of Figure \ref{fig1}d is displayed in black and gray to prevent any potential confusion regarding the concept of a band pair and its associated color representation.

\begin{figure*}[t!]
    \centering
    \includegraphics[width=0.9\textwidth]{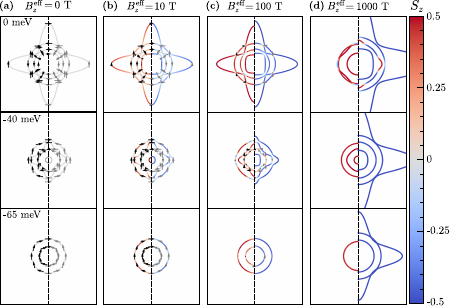}
    \caption{\textbf{Iso-energy lines and spin textures}. The contour energies are the same as in Fig.\ref{fig1} ($\SI{0}{\milli\electronvolt}$, $\SI{-40}{\milli\electronvolt}$, and $\SI{-65}{\milli\electronvolt}$ in that order). The exchange field strengths are different ($B^\text{eff}_z$ = $\SI{0}{\tesla}$ (a), $\SI{10}{\tesla}$ (b), $\SI{100}{\tesla}$ (c), and $\SI{1000}{\tesla}$ (d)). The arrows represent the in-plane spin expectation values, while the color indicates the out-of-plane spin expectation values. For better visibility, the left and right sides of each figure correspond to the higher energy band and lower energy band of a pair, respectively, except for $\SI{1000}{\tesla}$, where it corresponds to positive (resp. negative) values of $S_z$ at $\Gamma$.
}
    \label{fig2}
\end{figure*}

%As shown on Figure~\ref{fig1} the band-dependent magnetic field-induced Zeeman-like splitting modifies the density of states of the different bands at a given energy, giving rise to a more pronounced non-compensation of the Fermi contours with opposite helicities but also to a larger splitting in the region of the avoided crossings of particular efficiency for the Edelstein effects. Those two mechanisms should clearly boost the spin-charge interconversion. The counterpart is that the perpendicular magnetic field induces a reorientation of the spins out of plane, that should thwart spin-charge interconversion, thus competing with the former effect. 

To further highlight the impact of the exchange field, the spin textures at various field strengths, corresponding to the iso-energy contours depicted in Figure~\ref{fig1}, are presented in Figure~\ref{fig2}. When no out-of-plane exchange field is present (Fig. ~\ref{fig2}a), the spins are oriented within the plane, resulting in a zero out-of-plane spin density. The arrows in the figure, representing the in-plane spins, exhibit the familiar Rashba texture for circular contours, with the absolute values of the spin expectation values equal to $\hbar/2$. At higher energies, the spin texture becomes more intricate notably at avoided crossing points \cite{Vaz2019}.

As the magnetic exchange field is increased, the in-plane spin expectation values diminish notably in favor of increased out-of-plane spin expectation values giving rise to an out-of-plane equilibrium spin magnetization of the 2DEG. This change of orientation is particularly pronounced, with the absolute out-of-plane spin expectation values (represented by the color) close to $\hbar/2$ at $\SI{1000}{\tesla}$ (Fig. ~\ref{fig2}d). More importantly, this spin reorientation would lead to a reduction of the current-induced non-equilibrium in-plane spin polarization by the Edelstein effect and to a reduction of the produced charge current by the inverse Edelstein effect when spin current is injected. Figure~\ref{fig2} also evidences the large impact of the field-induced Zeeman-like splitting on the bands present at the different energies and the potential of this out-of-plane field to reinforce the contrast between contours with opposite spin chiralities and thus to boost the spin-charge interconversion. The effective impact of those two counteracting effects on spin-charge interconversion is described in the next section.  

\section{Edelstein effect}
The spin Edelstein effect, which is the main focus of this work, corresponds to a non-equilibrium spin density, leading to a finite magnetization, induced by an external electric field. In order to quantify this effect, we define the spin Edelstein susceptibility $\chi^s$, 
\begin{equation}\label{eq:susceptibility}
    \vec m = \vec m_0 + \chi^s \vec E
\end{equation}
with $\vec m$ the total magnetic moment per unit cell, $\vec m_0$ the equilibrium magnetic moment per unit cell, and $ \chi^s \vec E$ the magnetic moment originating from the current-induced spin density. The rank-2 tensor $\chi^s$ is the spin Edelstein susceptibility, and $\vec E$ is the external electric field (producing the current). However, in addition to this spin Edelstein effect (SEE), the electrons' orbital magnetic moments can also give rise to a finite current-induced magnetization, called orbital Edelstein effect (OEE)~\cite{Levitov_magnetoelectric_1985,Koretsune_magneto_2012,Yoda_current_2015,Go_toward_2017, Salemi_orbitally_2019,Hara_current_2020}. It has been shown that at STO interfaces, the OEE can exceed the SEE remarkably~\cite{johansson_spin_2021}. However, given that the experimental realization of the Edelstein effect typically observes only the SEE~\cite{Rojas_spin_2013, Rojas_spin_2016, Vaz2019, johansson_spin_2021}, our work focuses only on the SEE.

Within the semiclassical Boltzmann transport theory, the spin Edelstein susceptibility defined by Eq.~\eqref{eq:susceptibility} is given by
\begin{equation}\label{eq:suscept_Boltzmann}
\chi^s_{ij} = g_\text s \frac{A_0 e \mu_\text B}{A \hbar} \sum \limits_{\vec k} \tau_{\vec k} \vec S_{\vec k} ^i \vec v_{\vec k} ^j \delta \left( \epsilon_{\vec k} - \epsilon_\text F \right) \ .
\end{equation}
Here, $A_0$ is the area of the unit cell, $A$ is the area of the sample, $e$ is the absolute value of the elementary charge, the multi-index $\vec k$ represents the crystal momentum of an electronic state $\ket{\vec k}$ as well as the band index, $\tau_{\vec k}$ is the momentum relaxation time, which is set constant $\tau_{\vec k} = \tau_0$ in the following, $\vec S_{\vec k}$ is the spin expectation value of the state $\ket{\vec k}$, $\vec v_{\vec k} $ is the group velocity, $\epsilon_{\vec k}$ is the energy of the state $\ket{\vec k}$, and $\epsilon_\text F$ is the Fermi energy. 
We have used the relaxation time approximation to solve the linearized Boltzmann equation and assumed zero temperature. Details of the Boltzmann transport theory can be found in Appendix~\ref{sec:Boltzmann}.

The symmetry of the system allows for nonzero tensor elements $\chi_{xy}^s=-\chi_{yx}^s$, which quantify a current-induced magnetization oriented perpendicular to the applied electric field. In the presence of a nonzero magnetic exchange field, the current-induced magnetization can also exhibit a component parallel to $\vec E$  due to the broken time-reversal symmetry, represented by $\chi_{xx}^s= \chi_{yy}^s$. However, in the system under consideration, the extrinsic Edelstein effect, calculated within the Boltzmann approach, does not provide any contribution to $\chi_{xx}^s$. This tensor element can be nonzero only due to the intrinsic Edelstein effect, calculated for example using the Kubo approach.
As shown in Refs.\cite{Dyrdal_current_2017, Johansson_PhD}, for a pure Rashba system with Zeeman-like splitting these intrinsic contributions to the current-induced magnetization are several orders of magnitude smaller than the extrinsic contributions as well as the equilibrium magnetization. Therefore, they are not discussed in this work.

\begin{figure}%[t]
    \centering
    \includegraphics[width=\linewidth]{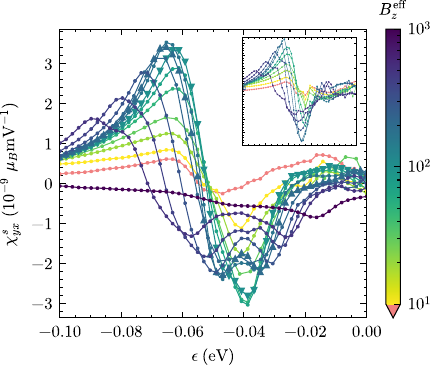}
    \caption{\textbf{Spin Edelstein effect conversion efficiency as a function of the energy at various exchange field strengths.} 
    The exchange field strengths range from $\SIrange{0}{1000}{\tesla}$ and the energy ranges from $\SIrange{-100}{0}{\milli\electronvolt}$. The conversion efficiency at zero field is represented in red. To compare with previous figures, the downward (resp. upward)-pointing triangles represent $\SI{100}{\tesla}$ (resp. $\SI{180}{\tesla}$). %The inset corresponds to the multiplication of the averaged sub-band splitting with the average modulus of the in-plane spin expectation value; for details see main text.
    The inset corresponds to the product of average wave vector $ k$ and the average modulus of the in-plane spin expectation value of each band; for details see main text.}
    \label{fig3}
\end{figure}

Figure~\ref{fig3} illustrates the SEE conversion efficiency as a function of the Fermi level at various exchange field strengths. The efficiency at $B_z^\text{eff}=\SI{0}{\tesla}$ is depicted in red. It exhibits two positive maxima at $\SI{-65}{\milli\electronvolt}$ and $\SI{-15}{\milli\electronvolt}$ and a negative maximum around $\SI{-50}{\milli\electronvolt}$. The introduction of an out-of-plane exchange field notably enhances this efficiency around $\SI{-65}{\milli\electronvolt}$, by approximately an order of magnitude at $\SI{180}{\tesla}$; beyond this value the efficiency gradually decreases. A similar trend is observed in the $\SIrange{-50}{-40}{\milli\electronvolt}$ region, albeit with a negative efficiency, but the maximum is attained at a lower field (approximately $\SI{70}{\tesla}$). These findings highlight the significant impact of the exchange field on the SEE conversion efficiency and provide crucial insights into its optimal operating conditions. By fine-tuning both $B_z^\text{eff}$ and $\epsilon_\text F$, a strong increase of the SEE in STO-based 2DEGs can be achieved.

%As shown in Figure~\ref{fig1} the band-dependent magnetic field-induced Zeeman-like splitting modifies the density of states of the different bands at a given energy, giving rise to a more pronounced non-compensation of the Fermi contours with opposite helicities but also to a larger splitting in the region of the avoided crossings of particular efficiency for the Edelstein effects. Those two mechanisms should clearly boost the spin-charge interconversion. The counterpart is that the perpendicular magnetic field induces a reorientation of the spins out of plane, that should thwart spin-charge interconversion, thus competing with the former effect. 

To understand the strong $B_z^\text{eff}$ dependence of the SEE, in particular the emergence of the maxima, we recall the Edelstein efficiency in an ordinary isotropic Rashba system~\cite{Edelstein_spin_1990}. For a single Rashba band pair, the Edelstein susceptibility scales with the band splitting in $\vec k$ space and the modulus of the $\vec k$ dependent in-plane spin expectation value. This relation is now transferred to the multi-band STO system.
The inset in Figure~\ref{fig3} presents a simplified  calculation based on the interplay between two competing effects: the sub-band splitting and the out-of-plane spin polarization. More specifically, to obtain this value, we take the energy-dependent average modulus of the wave vector $k$ (or radius for circular contours) of each band and multiply it by the average modulus of the $\vec k$ dependent in-plane spin expectation value. Next, we compute the difference between outer and inner band value of a pair. For fields higher than $\SI{180}{\tesla}$ where the concept of `pairs' is more challenging, we pair the highest energy band with the second highest energy band to form the first pair (1+2), and continue in this manner until reaching the lowest energy band (in the pair 7+8)~\footnote{Note that this classification of bands affects only the approximated curve (inset) but not the accurate calculation of the Edelstein effect (main figure) based on Eq. (4).}. Therefore we calculate $(k \cdot  s_{\text{in-plane}})_{\text{band i}} - (k \cdot s_\text{in-plane})_{\text{band i+1}}$. This approach yields a result that closely resembles the efficiency calculated within the full Boltzmann approach (Figure~\ref{fig3}), with a maximum value around $\SI{180}{\tesla}$. Hence, we conclude that the SEE provided by each band pair approximately scales with $(k \cdot  s_{\text{in-plane}})_{\text{band i}} - (k \cdot s_\text{in-plane})_{\text{band i+1}}$. In general, large sub-band splitting and large in-plane spin expectation values enhance the SEE. However, the multi-band character of the STO-based 2DEG as well as hybridization lead to a much more intricate energy dependence of the SEE compared to a trivial Rashba system, as discussed in detail in Ref.~\cite{Vaz2019} for $B_z^\text{eff}=\SI{0}{\tesla}$.

\begin{figure}
    \centering
    \includegraphics[width=\linewidth]{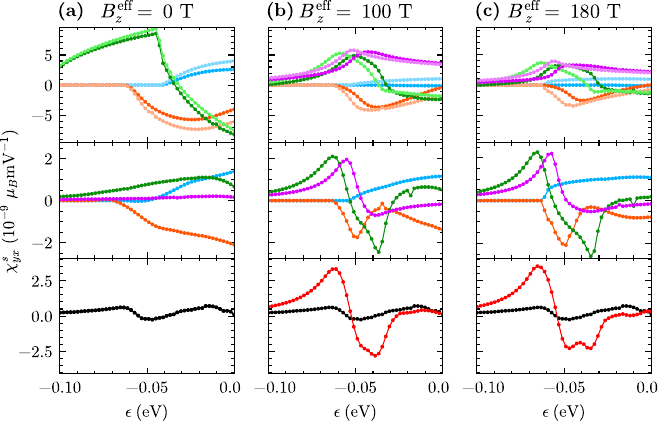}
    \caption{\textbf{Contributions to the SEE conversion efficiency at various exchange field strengths at $B_z^\text{eff}=\SI{0}{\tesla}$ (a), $B_z^\text{eff}=\SI{100}{\tesla}$ (b) and $B_z^\text{eff}=\SI{180}{\tesla}$ (c)}. The top panel represents the efficiency of each band (with the contribution of the inner band multiplied by $-1$ for better comparison). The middle panel shows the efficiency contribution of each band pair. The bottom panel displays the total efficiency. The total efficiency at $\SI{0}{\tesla}$ is represented in black.
}
    \label{fig4}
\end{figure}

Figure~\ref{fig4} provides a decomposition of the calculated SEE efficiency. In the top panel, we observe the efficiency contributed by each individual band whereas the middle panel shows the efficiency contribution of each band pair and the bottom panel displays the total efficiency. 

The energy dependence of the SEE is tied to the intricate band structure at the STO interface. At zero field and low energy, the $d_{xy}$ bands (magenta and green) exhibit notable spin-charge conversion efficiency per band (the magenta bands are above the top panel window). Nevertheless, the pair contribution remains modest, resulting in a relatively small value of $\chi_{yx}^s$. Upon increasing the energy, the first observed extremum corresponds to the onset at $\SI{-63}{\milli\electronvolt}$ of the first $d_{yz}$/$d_{xz}$ band (orange pair) with a contribution of opposite sign to $\chi_{yx}^s$. The SEE efficiency reaches a large negative value in the region of avoided crossing points between bands of $d_{xy}$ and $d_{yz}$/$d_{xz}$ character ($\SI{-50}{\milli\electronvolt}$ to $\SI{-40}{\milli\electronvolt}$, along the $\left\langle 1 0 0 \right \rangle$ directions). The negative maximum of the efficiency is then reached at the onset ($\SI{-63}{\milli\electronvolt}$) of the blue band, with positive contribution.

The situation becomes more intricate in the presence of an exchange field, as the contribution of each band pair displays a less monotonic behavior, with pronounced positive and negative maxima. The general trend is however that, whereas the SEE efficiency value of each individual band diminishes with increasing exchange field strength because of the reduced in-plane spin expectation values, excitingly, the contributions of the band pairs and the total SEE actually increase with the field up to $\SI{180}{\tesla}$ due to Zeeman-like splitting both in energy and $\vec k$. 

Below the onset of the orange band, the energy dependence of the SEE is not changed qualitatively by the $\SI{100}{\tesla}$ ($\SI{180}{\tesla}$) exchange fields. The avoided crossing between the orange and green bands occurs around $\SI{-50}{\milli\electronvolt}$ ($\SI{-60}{\milli\electronvolt}$), whereas the avoided crossing between green and magenta band pairs is around $\SI{-45}{\milli\electronvolt}$ ($\SI{-50}{\milli\electronvolt}$). This region eventually coincides with the onset of the lowest blue band as they are shifted by the field, altering the negative maximum.

\begin{figure}
    \centering
    \includegraphics[width=\linewidth]{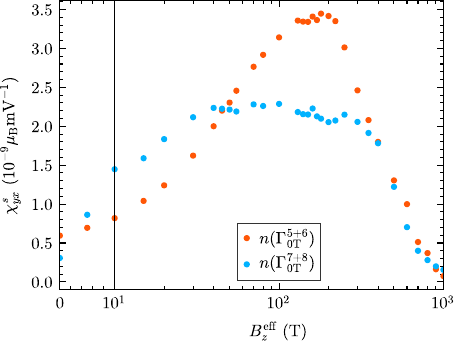}
    \caption{\textbf{Spin Edelstein effect conversion efficiency as a function of exchange field at constant charge carrier density}. The SEE conversion efficiency is calculated at a fixed carrier density corresponding to the edge of the orange band (orange dots, $n = 5.44 \times 10^{13} \mathrm{cm}^{-2}$ at $\epsilon = \SI{-63}{\milli\electronvolt}$) and the blue band (blue dots, $n = 8.99 \times 10^{13} \mathrm{cm}^{-2}$ at $\epsilon = \SI{-43}{\milli\electronvolt}$) at $\SI{0}{\tesla}$, representing the positive and negative maxima of the efficiency at $\SI{0}{\tesla}$. The scale of the abscissa is linear from $\SI{0}{\tesla}$ to $\SI{10}{\tesla}$ and semi-logarithmic between $\SI{10}{\tesla}$ and $\SI{1000}{\tesla}$.
}
    \label{fig5}
\end{figure}

To further illustrate the enhancement of the efficiency due to the exchange field, we  present in Fig.~\ref{fig5} the SEE efficiency as a function of the exchange field strength, at a constant charge carrier density \footnote{Note that the shift of the bands caused by the exchange field alters the density of states and therefore effectively shifts the location of the Fermi energy. To compare the Edelstein signals for different field strengths it is more significant to compare configurations with equal carrier density than equal energy.}. As explained in Figures~\ref{fig1} and \ref{fig4}, the edge of the orange band pair, which is the onset of the first heavy bands, remains unchanged in energy. Moreover, it coincides with the maximum efficiency, as observed in Figure~\ref{fig3}. Therefore, we selected the carrier density at this energy (without exchange field; $n= \SI{5.44e13}{\per\centi\meter\tothe{2}}$ at $\epsilon = \SI{-63}{\milli\electronvolt}$) as a reference for all field strengths. 

The negative maximum of $\chi_{yx}^s$ observed around $\SI{-40}{\milli\electronvolt}$ in Figure~\ref{fig3} is loosely related to the edge of the blue band pair. Contrary to the orange band, its exact energy position varies with $B_z^\text{eff}$. The blue dots in Fig.~\ref{fig5} represent the Edelstein susceptibility for a charge carrier density of $n= \SI{8.99e13}{\per\centi\meter\tothe{2}}$, which corresponds to the edge of the blue band at $B_z^\text{eff}=0$ ($\epsilon = \SI{-43}{\milli\electronvolt}$).

The presence of an out-of-plane magnetic exchange field generally enhances the splitting of each band pair in terms of wave vector ($k$). It also tilts the spin expectation values out of the plane, thereby reducing their in-plane components. Regarding the Edelstein conversion efficiency, these two effects compete with each other, resulting in the non-monotonic behavior of the spin Edelstein effect as a function of $B_z^\text{eff}$.

For the carrier density corresponding to the edge of the orange band pair, the maximum of the curve is reached at an exchange field of $\SI{190}{\tesla}$. This finding aligns with the previous observations. At $\SI{180}{\tesla}$, as we can see in Fig.~\ref{fig1}, the lower energy blue band is about to cross the orange bands. Slightly above this exchange field, they overlap, which modifies both the amplitude and the position of the positive maximum of the efficiency. Notably, this maximum efficiency is nearly one order of magnitude larger than the maximum efficiency without a field. On the other hand, for the blue band, it appears that we reach a plateau instead of a distinct peak. As emphasized in Figure~\ref{fig4}, the negative maximum does not correspond to a well-defined peak.

\section{Conclusion}
In conclusion, this study sheds light on the influence of a perpendicular exchange field on the spin-charge interconversion in the Rashba 2D electron gas at SrTiO$_3$ interfaces. 
Remarkably, increasing the magnetic field yields an up to six-fold increase of the spin-charge interconversion efficiency which we explain by the competing exchange and Rashba interactions. We have found that an out-of-plane magnetic field produces a Zeeman splitting of the band pairs thereby increasing the Edelstein effect. At the same time, it strongly modifies the Rashba-type spin textures and tilts them from their original in-plane direction towards out-of-plane thereby reducing the Edelstein effect. Therefore, a trade-off of the two competing effects causes a maximum which we observe close to a field strength of $\SI{190}{\tesla}$ which is where shifted bands overlap close to the $\Gamma$ point. 

% Original: We have found that, in addition to producing a Zeeman splitting of the band pairs, an out-of-plane magnetic field strongly modifies the Rashba-split spin textures and tilts them from their original in-plane direction towards out-of-plane. Remarkably, increasing the magnetic field yields a six-fold increase of the spin-charge interconversion. After reaching this maximum value at a field strength of about $\SI{190}{\tesla}$, the shifted bands overlap and the efficiency starts to decrease. This efficiency enhancement stems from the induced Zeeman-like splitting of the band pairs both in energy and wave vector, especially for exchange field in the $\SIrange{100}{200}{\tesla}$ range.

Our findings open new routes to provide more efficient SOC materials or interfaces for emerging devices such as the MESO device proposed by Intel \cite{manipatruni_scalable_2019}. They are also relevant because several studies have reported STO 2DEGs at interfaces with a magnetic oxide such as EuTiO$_3$ \cite{stornaiuolo_tunable_2016}, EuO \cite{lomker_two-dimensional_2017}, or LSMO \cite{lu_long-range_2016}, with indications of induced magnetism in the 2DEG \cite{stornaiuolo_tunable_2016,lu_long-range_2016,brehin_coexistence_2023}. Finally, our results suggest a general approach that could be tested on other Rashba systems endowed with magnetic interactions, such as PdCoO$_2$ interface/surface states\cite{lee_nonreciprocal_2021}, EuO/KTaO$_3$ \cite{zhang_high-mobility_2018} or certain van der Waals heterostructures \cite{shi_recent_2023}.

\section*{Acknowledgement}
This work received support from the ERC AdG ``FRESCO" ($\#$833973). I.M. acknowledges support from the DFG under SFB TRR 227. 

\appendix
\section{Tight-binding model for STO-based 2DEGs}\label{sec:model_Hamiltonian}
Following Refs.~\cite{khalsa_theory_2013, Zhong2013, Vivek2017, Vaz2019, johansson_spin_2021}, we describe the electronic states which are relevant for the formation of the 2DEG at STO interfaces by an effective tight-binding Hamiltonian $\mathcal H_\text{STO}$ in the basis set $\left(d_{xy \uparrow}^{(1)},d_{xy \uparrow}^{(2)}, d_{yz \uparrow}, d_{zx \uparrow}, d_{xy \downarrow}^{(1)},d_{xy \downarrow}^{(2)}, d_{yz \downarrow}, d_{zx \downarrow}  \right) $. Two $d_{xy}$ orbitals occur due to the confinement of the 2DEG along the $z$ direction. The full Hamiltonian $\mathcal H_\text{STO}$ can be split into three terms,
\begin{equation}\label{eq:Hamiltonian_STO} 
\mathcal H_\text{STO} = \mathcal H_0 + \mathcal H_\lambda + \mathcal H_\text{OM} \ .
\end{equation}
Here, $\mathcal H_0$ is the free-electron-like Hamiltonian neglecting spin-orbit coupling,
\begin{equation} \label{eq:Hamiltonian_0}
    \mathcal H _0 = \mathds{1} \otimes \left(  \begin{array}{c c c c}
     \epsilon_{xy}^{(1)}  & 0 & 0   & 0 \\
       0  & \epsilon_{xy}^{(2)} & 0 & 0 \\
       0 & 0 & \epsilon_{yz} & 0 \\
       0 & 0 & 0& \epsilon_{zx} 
    \end{array} \right)
\end{equation}
with 
\begin{equation}
    \begin{split}
        \epsilon_{xy}^{(i)}& = 2 t \left( 2 - \cos a k_x - \cos a k_y \right) + \epsilon_{xy 0}^{(i)} \ , \ \  i=1,2 \\ 
        \epsilon_{yz} & = 2 t \left( 1 - \cos a k_y \right) + 2 t_h \left( 1 - \cos a k_x  \right) + \epsilon_{z0} \  , \\ 
        \epsilon_{zx} & = 2 t \left( 1 - \cos a k_x \right) + 2 t_h \left( 1 - \cos  a k_y  \right) + \epsilon_{z0} \ . 
    \end{split}
\end{equation}
Here, $\vec{k}$ is the crystal momentum, $a$ is the lattice constant, the parameters $t$ and $t_h$ describe nearest-neighbor hopping of the light and heavy bands, respectively, and $\epsilon_{xy0/z0}$ correspond to the on-site potentials.

The second term of the right-hand side of Eq.~\eqref{eq:Hamiltonian_STO} corresponds to atomic spin-orbit coupling,
\begin{equation}
 \mathcal H_\lambda = \frac{2}{\hbar ^2} \lambda \vec{\mathcal{L}} \cdot \vec{\mathcal{S}} \ ,
\end{equation}
with $\vec{\mathcal L }$ and $\vec{\mathcal S}$ the orbital angular momentum and spin operators, respectively, which are represented in our particular basis set as
\begin{equation}\label{eq:L_operator}
    \mathcal L_i = \hbar  \mathds{1} \otimes  l_i  \ , \ \ i=x,y,z
\end{equation}
with
\begin{equation}\label{eq:li}
    \begin{split}
         l_x &= \left( \begin{array}{ c c c c}
         0 & 0 & 0  & - i   \\
         0 &  0 & 0 & -i \\
         0 & 0 & 0 & 0 \\
         i & i & 0 & 0 
        \end{array} \right) \ , \\
         l_y &= \left( \begin{array}{ c c c c}
         0 & 0 & i  & 0   \\
         0 &  0 & i & 0 \\
         -i & -i & 0 & 0 \\
         0 & 0 & 0 & 0 
        \end{array} \right) \ , \\
         l_z & = \left( \begin{array}{ c c c c}
         0 & 0 & 0  & 0  \\
         0 &  0 & 0 & 0 \\
         0 & 0 & 0 & i \\
         0 & 0 & -i & 0 
        \end{array} \right) \ ,
    \end{split}
\end{equation}
 and
 \begin{equation}\label{eq:S_operator}
    \mathcal S_i = \frac{\hbar}{2}  \sigma_i \otimes \mathds{1}  
 \end{equation}
 with $ \sigma_i$ the Pauli spin matrices. Finally, the term $\mathcal H_\text{OM}$ in Eq.~\eqref{eq:Hamiltonian_STO} corresponds to inter-atomic orbital mixing, arising from the broken inversion symmetry at the interface leading to a deformation of the orbitals~\cite{Zhong2013, khalsa_theory_2013, Petersen2000},
 \begin{widetext}
 \begin{equation}\label{eq:H_OM}
     \mathcal H_\text{OM} = \mathds{1} \otimes \left( \begin{array}{cccc}
       0 & 0  & 2 i g_1 \sin a k_x & 2 i g_1 \sin a k_y  \\
       0 & 0 & 2 i g_2 \sin a k_x & 2 i g_2 \sin a k_y \\ 
       -2 i g_1 \sin a k_x & -2 i g_2 \sin a k_x & 0 & 0 \\
       -2 i g_1 \sin a k_y & -2  i g_2 \sin a k_y & 0 & 0 
     \end{array} \right) \ .
 \end{equation}
 \end{widetext}
 In this work, we use the following parameters, adopted from Refs.~\cite{Vivek2017, Vaz2019},
 \begin{equation}\label{eq:model_parameters}
     \begin{split}
         \epsilon_{xy0}^{(1)} & = - \SI{205}{\milli\electronvolt} \ , \ \ \ \  t= \SI{388}{\milli\electronvolt} \ , \\
          \epsilon_{xy0}^{(2)} & = - \SI{105}{\milli\electronvolt} \ , \ \ \   t_h= \SI{31}{\milli\electronvolt} \ , \\
           \epsilon_{z0} & = - \SI{54}{\milli\electronvolt} \ , \ \ \ \ \  g_1= \SI{2}{\milli\electronvolt} \ , \\
            \lambda & = - \SI{8.3}{\milli\electronvolt} \ , \ \ \ \  g_2= \SI{5}{\milli\electronvolt} \ .
     \end{split}
 \end{equation}
%\section{Rashba model with Zeeman term}
%STO-based 2DEGs exhibit Rashba-like states, meaning that close to the $\Gamma$ point, each individual band pair can be approximated by an effective two-band Rashba model~\cite{Bychkov1984oscillatory, Bychkov1984properties, Rashba1960}. 
%\begin{equation}\label{eq:Rashba_model}
%    \mathcal H_\text R = \frac{\hbar ^2 k^2}{2 m} + \alpha_\text R \left( k_x \hat \sigma_y - k_y \hat \sigma x \right) 
%\end{equation}
%with $\hbar$ the reduced Planck constant, $\vec k$ the crystal momentum, $m$ the effective mass, $\alpha_\text R$ the effective Rashba parameter, and $\hat \vec \sigma$ the Pauli spin vector acting on the spin degree of freedom. The basis of the two-band Hamiltonian is $\left(\Ket{\uparrow}, \Ket{\downarrow}\right)$. Neglecting any orbital character of the electronic states, an external magnetic field $\vec B$ couples to the electrons' spin via an additional Zeeman term in the Hamiltonian,
%\begin{equation}
%    \label{eq:Zeeman_Rashba}
%    \mathcal H _ \text Z = \frac{\mu_\text B}{\hbar} g_s \vec S \cdot \vec B
%\end{equation}
%with $\mu_\text B$ the Bohr magneton, $g_s$ the electon Landé factor, and $\vec S $ the spin operator with $\vec S = \frac{\hbar}{2} \hat \vec \sigma$ in the two-band basis.
%Similarly, the influence of a finite magnetization can be expressed via the exchange coupling Hamiltonian,
%\begin{equation}
%   \label{eq:Rashba_exchange} 
%   \mathcal H_\text{ex}= - 
%\end{equation}
\section{Boltzmann transport theory}\label{sec:Boltzmann}
Within the semiclassical Boltzmann transport theory, the current-induced magnetic moment per unit cell originating from the spin Edelstein effect is given by
\begin{equation}\label{eq:magnetic_moment_Boltzmann}
\vec m = - \frac{ A_0 g_s \mu_\text B}{A \hbar} \sum \limits_{\vec k} f_{\vec k} \vec S _{\vec k} \ ,
\end{equation}
with $A_0$ the area of the unit cell, $A$ the area of the sample, and $f_{\vec k}$ the distribution function, which is split into an equilibrium part, the Fermi-Dirac distribution function $f_{\vec k}^0$, and a nonequlibrium part $g_{\vec k}$.

In magnetic systems, the term of Eq.~\eqref{eq:magnetic_moment_Boltzmann} containing $f_{\vec k}^0$ gives rise to an equilibrium magnetization $\vec m_0$. In the STO-based 2DEG discussed in this work, the magnetic exchange field $B_\text{eff}^z$ induces a finite out-of-plane equilibrium magnetization. The application of an external electric field $\vec E$ leads to a change of the distribution function, represented by $g_{\vec k}$, whose contribution to Eq.~\eqref{eq:magnetic_moment_Boltzmann} is a current-induced nonequilibrium magnetic moment, the Edelstein effect.

The nonequilibrium distribution function $g_{\vec k}$ is determined by solving the Boltzmann equation. Here, we consider a spatially homogeneous and stationary system, 
\begin{equation}\label{eq:Boltzmann}
\dot{\vec k} \frac{\partial f_{\vec k}}{\partial \vec k}=\left(\frac{\partial f_{\vec k}}{\partial t} \right) _\text{scatt}.
\end{equation}
In the presence of an external electric field, the semiclassical equation of motion reads
\begin{equation}\label{semiclassical_k}
\dot{\vec k}= - \frac{e}{\hbar} \vec E .
\end{equation}
Within the relaxation time approximation, the scattering-term is expressed by
\begin{equation}\label{eq:RTA}
    \left(\frac{\partial f_{\vec k}}{\partial t} \right) _\text{scatt} = - \frac{1}{\tau_{\vec k}}g_{\vec k}
\end{equation}
with $\tau_{\vec k}$ the relaxation time, which is assumed constant, $\tau_{\vec k}=\tau_0$ in our calculations.

The Boltzmann equation~\eqref{eq:Boltzmann} is then solved by 
\begin{equation}\label{eq:Boltzmann_solution}
    f_{\vec k} = f_{\vec k }^0 + \frac{\partial f_{\vec k}}{\partial \epsilon} e \tau_0 \vec v_{\vec k} \cdot \vec E \ .
\end{equation}
Inserting this solution into Eq.~\eqref{eq:magnetic_moment_Boltzmann} and assuming zero temperature, Eq.~\eqref{eq:susceptibility} for the Edelstein susceptibility, characterizing the nonequilibrium current-induced magnetic moment, is obtained.
\newpage

%\bibliography{references}% Produces the bibliography via BibTeX.

%apsrev4-2.bst 2019-01-14 (MD) hand-edited version of apsrev4-1.bst
%Control: key (0)
%Control: author (8) initials jnrlst
%Control: editor formatted (1) identically to author
%Control: production of article title (0) allowed
%Control: page (0) single
%Control: year (1) truncated
%Control: production of eprint (0) enabled
\providecommand{\noopsort}[1]{}\providecommand{\singleletter}[1]{#1}%

\end{document}